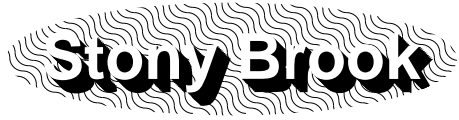

# SUPERSPACE DUALITY IN LOW-ENERGY SUPERSTRINGS

W. Siegel[1]

*Institute for Theoretical Physics*
*State University of New York, Stony Brook, NY 11794-3840*

**ABSTRACT**

We extend spacetime duality to superspace, including fermions in the low-energy limits of superstrings. The tangent space is a curved, extended superspace. The geometry is based on an enlarged coordinate space where the vanishing of the d'Alembertian is as fundamental as the vanishing of the curl of a gradient.

---

[1] Work supported by National Science Foundation grant PHY 9211367.
Internet address: siegel@max.physics.sunysb.edu.

# 1. INTRODUCTION

In a previous paper [1], we showed how spacetime duality invariance [2] (a symmetry which mixes graviton and axion) of the bosonic sectors of strings, which is preserved in low-energy limits, can be made manifest by using a new formalism for axionic gravity. We will first present that formalism in a yet more general way to allow it to be applied to fermions as well. The basic idea is to introduce a generalized vierbein with a local tangent-space gauge invariance involving general linear groups rather than the usual orthogonal (Lorentz) groups, a reinterpretation of Cartan's approach to gravity. There are independent left- and right-handed GL groups corresponding to the left- and right-handed modes of the string, so the field theories describing the low-energy behavior of strings still have the string's handedness built in. Possible advantages of this string-inspired approach to spacetime geometry include: new gauges that make the field theory Feynman rules more closely resemble those from string theory and allow solutions related by duality to be found more directly, more convenient insertions of background fields for $\sigma$-model calculations, and better understanding of (super)string field theory.

In the following section we introduce the background fields for the hamiltonian formalism of strings. The oscillators $Z_M$ of the string carry a superindex $M$; duality is just a global SO(d,d+n) transformation acting on that index. The commutation relations of these oscillators involve a constant metric $\eta_{MN}$; the SO(d,d+n) duality invariance is a subgroup of the OSp(D,D+n|2D$'$) symmetry (SO(D,D+n) in the bosonic case) that preserves this metric. Gauge transformations $\lambda^M$ for massless fields are generated by $Z_M$. Since $Z_M$ form an affine Lie algebra (and not an ordinary Lie algebra), the algebra of this group defines a new Lie derivative. The zero-modes of $Z_M$ are the derivatives $\partial_M$, which include not only the usual momenta but also winding numbers. Although the winding numbers vanish on the massless background fields we consider, this condition is imposed in calculations only as $\partial^M \partial_M = 0$. This is not a mass-shell condition (the metric $\eta_{MN}$ is off-diagonal), but a partner to the usual relation $\partial_{[M}\partial_{N)} = 0$ for partial derivatives.

Background fields are introduced by covariantizing $Z_M$ as $\Pi_A = e_A{}^M Z_M$ with the background vielbein $e_A{}^M$. Its transformation under $\lambda^M$ is a cross between a covariant vector and a contravariant one, a consequence of the modification of the Lie derivative. The tangent-space superindex $A$ separately labels left- and right-handed oscillators. It defines a tangent-space metric $g_{AB}$ that is not constant; as a result, the local tangent-space symmetry involves general linear groups rather than orthogonal groups. To allow definition of the left- and right-handed conformal generators (the first-class constraints), this metric is constrained to have its left-right off-diagonal part vanish, and the tangent-space gauge group is constrained to not mix left and right indices. The dilaton field is required (as a density) to define actions.

In section 3 we discuss the supersymmetric case. Additional constraints are required on the tangent-space metric and gauge group to preserve the second-class constraints. These constraints allow the existence of connection-independent torsions, as in ordinary superspace for pure supergravity.

Unlike ordinary superspace, all these torsions are found to be trivial when we consider constraints on them in section 4. The constraints are those that preserve the form of the algebra of first- and second-class constraints. We also discuss further restrictions on the allowed tangent-space transformations of the second-class constraints; these are necessary to define representation-preserving constraints, such as those that allow the existence of chiral superfields, like the dilaton.

Section 5 defines the conventional gauge, where the tangent-space symmetry is fixed enough to allow identification of the surviving components of the vielbein with the usual (super)graviton and axion (super)fields. In this gauge the torsion constraints take the familiar form. The usual super Yang-Mills spinor field strength appears as a component of the vielbein.

At least in three and four spacetime dimensions, the torsion constraints can be solved off shell. In D=3 the solution picks certain parts of the vielbein as independent. In section 6 we find the D=4 solution in terms of prepotentials, which appear as in usual superspace as exponential operators, in the form of complexified group elements. Duality then acts in the obvious way on the index of the prepotential $W^M$.

For a deeper understanding of the geometry of this superspace we consider covariant derivatives in section 7. Because of the form of the new Lie bracket, not all connections can be determined, so only certain types of covariant differentiation are allowed. However, the treatment of $\partial^M \partial_M = 0$ on equal footing with $[\partial_M, \partial_N\} = 0$ means new torsions and curvatures can be defined whose construction involves traces on an equal footing with antisymmetrization. The use of traces is related to the existence of the dilaton as the integration measure.

In section 8 we construct the curvature tensors and actions for the bosonic case. The field equations are the generalization of the Ricci tensor and scalar. The lagrangian is simply this scalar, with the dilaton as the measure. The index structure of this tensor, even in the supersymmetric case, explicitly reflects the string property that closed string states are the direct product of left-handed open-string states with right-handed ones.

In the final section we review the basic results and their equations.

## 2. ALGEBRAS AND SYMMETRIES

We begin in the hamiltonian formalism by considering the affine Lie algebras of general string creation and annihilation operators (in the $\sigma$ representation), and the Virasoro algebras in terms of them, in the presence of background fields. The "bare" algebra is expressed in terms of a basis of oscillators $Z_M$ with "super"index $M = (_\mathbf{M}, {}^\mathbf{M}, \hat{m})$, where $\mathbf{M}$ is the index carried by the (graded) coordinates $x^\mathbf{M}$ of the background fields, and $\hat{m}$ labels additional directions with no corresponding coordinates, like the 16 compactified dimensions of the right-handed sector of the heterotic string:

$$[Z_M(1), Z_N(2)\} = i\delta'(2-1)\eta_{MN}$$

$$Z_M = (P_\mathbf{M}, X'^\mathbf{M}, P_{\hat{m}} - \tfrac{1}{2}X'_{\hat{m}}) \quad \Rightarrow \quad \eta_{MN} = \begin{pmatrix} 0 & \delta_\mathbf{M}{}^\mathbf{N} & 0 \\ \delta^\mathbf{M}{}_\mathbf{N} & 0 & 0 \\ 0 & 0 & -\delta_{\hat{m}\hat{n}} \end{pmatrix}$$

$\delta_\mathbf{M}{}^\mathbf{N}$ is the usual Kronecker delta, while in $\delta^\mathbf{M}{}_\mathbf{N}$, for fermionic indices $\delta^\mu{}_\nu = -\delta_\nu{}^\mu$ from index reordering. $X(1)$ is the string coordinate and $P(1)$ its momentum, where "1" labels the $\sigma$ parameter.

We next define a generalization of general coordinate transformations based on this algebra:

$$\Lambda = \int d1 \; \lambda^M(X^\mathbf{N}(1))Z_M(1)$$

where $\lambda^M$ includes general (super)coordinate transformations $\lambda^\mathbf{M}$, axion (2-form) gauge transformations $\lambda_\mathbf{M}$, and abelian vector(-multiplet) gauge transformations $\lambda^{\hat{m}}$. There is also the gauge invariance of the gauge invariance

$$\delta\lambda^M = \partial^M \lambda, \quad \partial_M = (\partial_\mathbf{M}, \partial^\mathbf{M}, \partial_{\hat{m}}) \quad \Rightarrow \quad \delta\Lambda = \int Z^M \partial_M \lambda = \int \lambda' = 0$$

(We freely raise and lower indices with the metric $\eta_{MN}$.) Although the operators $\partial^\mathbf{M}$ and $\partial_{\hat{m}}$ (related to winding numbers in the full string theory) vanish on the gauge parameters and backgrounds we consider, this condition will be used explicitly only as the consequences $A' = Z^M \partial_M A$ and $(\partial^M A)(\partial_M B) = \partial^M \partial_M A = 0$ on arbitrary functions $A$, which states that all supermomenta are orthogonal, even to themselves.

Otherwise, we treat $\partial^{\mathbf{M}}$ and $\partial_{\hat{m}}$ as spacetime central charges under which the background fields are neutral (but see the discussion of duality later in this section). The commutation relations of this symmetry group (with group elements $e^{-i\Lambda}$) defines a new Lie derivative

$$[\Lambda_1, \Lambda_2] = -i\Lambda_{[1,2]} \quad \Rightarrow \quad \lambda_{[1,2]}^M = \lambda_{[1}^N \partial_N \lambda_{2]}^M - \tfrac{1}{2}\lambda_{[1}^N \partial^M \lambda_{2]N}$$

which preserves the gauge invariance of the gauge invariance

$$\delta \lambda_i^M = \partial^M \lambda_i \quad \Rightarrow \quad \lambda_{[1,2]} = \tfrac{1}{2}\lambda_{[1}^M \partial_M \lambda_{2]}$$

(Sign factors due to ordering of superindices are implicit.)

Duality is basically the global symmetry on the superindex $M$ that preserves the metric $\eta_{MN}$. Since $X'$ is part of $Z$, this symmetry holds only when considering states with no dependence on d of the bosonic components of $X$ (as for the n components of $X_{\hat{m}}$), allowing an SO(d,d+n) symmetry. However, since d can be arbitrary, this generally means performing manipulations that treat these indices as if there were the full OSp(D,D+n|2D$'$) symmetry. (Configurations that are independent of some of the D$'$ fermionic components of $X$ are probably not useful, since supersymmetry spinor derivatives have terms proportional to $\theta^\mu$ times components of the vierbein that are nonvanishing in nonsingular coordinate systems.)

We then write the affine Lie algebra $\Pi_A$ of left- and right-handed oscillators in background fields collectively by labeling them with a superindex $A = (\mathcal{A}, \tilde{\mathcal{A}})$, where both the left-handed index $\mathcal{A}$ and the right-handed index $\tilde{\mathcal{A}}$ can range over commuting as well as anticommuting values (with different ranges for left and right). The backgrounds fields are represented by the vielbein $e_A{}^M(x^\mathbf{N})$, which is an invertible square matrix, and can be used to define a curved tangent-space metric $g_{AB}$ analogously to ordinary gravity, in addition to the orthosymplectic coordinate-space metric $\eta_{MN}$ ($\eta_{[MN)} = 0$):

$$\Pi_A(1) = e_A{}^M(X^\mathbf{N}(1))Z_M(1), \quad g_{AB} = e_A{}^M e_B{}^N \eta_{MN}$$

$g_{AB}$ and $\eta_{MN}$ and their inverses are used to raise and lower their corresponding indices, but $g_{AB}$ is a function while $\eta_{MN}$ is a constant, the reverse of the usual situation. The commutation relations of this affine Lie algebra are:

$$[\Pi_A(1), \Pi_B(2)\} = i\delta'(2-1)\tfrac{1}{2}[g_{AB}(1) + g_{AB}(2)] + i\delta F_{AB}{}^C \Pi_C$$

$$F_{AB}{}^C \equiv \tfrac{1}{2}(f_{[ABD)} + e_{[A}g_{B)D})g^{DC} = f_{[AB)}{}^C + \tfrac{1}{2}f^C{}_{[AB)}$$

$$f_{ABC} \equiv (e_A e_B{}^M) e_{CM}, \quad e_A \equiv e_A{}^M \partial_M$$

$$\Rightarrow \quad f_{A(BC]} = e_A g_{BC}, \quad F_{A(BC]} = e_A g_{BC} - \tfrac{1}{2} e_{(B} g_{C]A}$$

We have used $A(1)\delta'(2-1) = A(2)\delta'(2-1) + \delta A'$. (On indices, "( ]" and "[ )" are graded symmetrization and antisymmetrization.) Since $(\partial^M A)(\partial_M B) = (e^A A)(e_A B) = 0$, we find

$$[e_A, e_B\} = F_{AB}{}^C e_C$$

The off-diagonal part of the metric is constrained to vanish so that we can define decoupling left- and right-handed Virasoro operators:

$$g_{\mathcal{A}\tilde{\mathcal{B}}} = 0$$

$$\Rightarrow \quad L_+ = \tfrac{1}{2} g^{\mathcal{A}\mathcal{B}} \Pi_{\mathcal{B}} \Pi_{\mathcal{A}}, \quad L_- = \tfrac{1}{2} g^{\tilde{\mathcal{A}}\tilde{\mathcal{B}}} \Pi_{\tilde{\mathcal{B}}} \Pi_{\tilde{\mathcal{A}}}; \quad L_+ + L_- = \tfrac{1}{2} Z^M Z_M$$

The gauge transformations are defined as

$$\delta \Pi_A = [-i\Lambda, \Pi_A] + \lambda_A{}^B \Pi_B, \quad \lambda_{\mathcal{A}}{}^{\tilde{\mathcal{B}}} = \lambda_{\tilde{\mathcal{A}}}{}^{\mathcal{B}} = 0$$

$$\Rightarrow \quad \delta e_A{}^M = (\lambda^N \partial_N e_A{}^M + e_{AN} \partial^{[M} \lambda^{N)}) + \lambda_A{}^B e_B{}^M, \quad \delta g_{AB} = \lambda^M \partial_M g_{AB} + \lambda_{(AB]}$$

where $\lambda_A{}^B$ describes left- and right-handed local (graded) GL invariances on the $\mathcal{A}$ and $\tilde{\mathcal{A}}$ indices. Under the $\lambda^M$ gauge transformations,

$$\delta f_{ABC} = (covariant) - e_A{}^M e_B{}^N e_C{}^P \partial_M \partial_{[N} \lambda_{P)}$$

Thus, of all of $f$, only $f_{[ABC)}$ and $f_{A(BC]} = e_A g_{BC}$ are covariant with respect to these transformations. Under the tangent-space transformations,

$$\delta f_{ABC} = (covariant) + (e_A \lambda_B{}^D) g_{DC}$$

$F_{AB}{}^C$ is a linear combination of $f_{[ABC)}$ and $e_A g_{BC}$, and so is $\lambda^M$-covariant. Its tangent-space transformation is

$$\delta F_{AB}{}^C = (covariant) + e_{[A} \lambda_{B)}{}^C + \tfrac{1}{2} (e^C \lambda_{[A|}{}^D) g_{D|B)}$$

Since not only $\partial_{[M} \partial_{N)} = 0$ (as usual) but also $\partial^M \partial_M = 0$, the $F$'s satisfy Bianchi identities for both curl and divergence:

$$\tfrac{1}{6} e_{[A} F_{BCD)} = \tfrac{1}{8} F_{[AB}{}^E F_{CD)E}, \quad (F_{AB}{}^C \overleftarrow{e}_C) = -1 \cdot [\overleftarrow{e}_A, \overleftarrow{e}_B\} = e_{[A}(1 \cdot \overleftarrow{e}_{B)})$$

where $1 \cdot \overleftarrow{e}_A \equiv e_A{}^M \overleftarrow{\partial}_M = (\partial_M e_A{}^M)$. The form of the first identity that follows directly from the II Jacobi identities is

$$e_{[A} F_{BC)}{}^D - \tfrac{1}{3} e^D F_{[ABC)} - F_{[AB|}{}^E F_{E|C)}{}^D + \tfrac{1}{2} F_{[AB|}{}^E e^D g_{E|C)} = 0$$

which resembles the usual Jacobi identities for structure functions.

Duality transformations also can be derived by considering $\lambda^M$ transformations for which $e_A{}^M$ is independent of d components $x^i$ of $x^m$. This only restricts $\lambda^M$ to be linear in $x^i$. Furthermore, $\lambda^M$ can be considered linear in coordinates $x^{\tilde{i}}$ and $x^{\hat{m}}$, corresponding to $\partial_{\tilde{m}}$ and $\partial_{\hat{m}}$, which we otherwise assume to vanish (on $e_A{}^M$). The net result is that in the $\lambda^M$ transformations we replace $\lambda^M \partial_M$ by 0 and $\partial_{[M} \lambda_{N)}$ by a constant matrix, with nonvanishing components for indices $(i, \tilde{i}, \hat{m})$, representing the global group SO(d,d+n).

Besides the vielbein there is also the dilaton, a scalar density needed for constructing actions ($det\ e_A{}^M = (det\ g_{AB})^{1/2}$ is not a density):

$$\delta \Phi^2 = \partial_M(\lambda^M \Phi^2)$$

This transformation and the one for the vielbein preserve the gauge invariance of the gauge invariance, since $\lambda_M$ appears only as $\lambda^M \partial_M$, $\partial_M \lambda^M$, and $\partial_{[M} \lambda_{N)}$. In treating the bosonic sectors of strings, $\Phi^2$ is just the square of a single scalar $\Phi$, but for the supersymmetric cases $\Phi^2$ is the square of the magnitude of a D$-$2 component scalar $\Phi_i$ (at least for D=3,4,6); the D$-$3 angular parts parametrize the coset space SO(D$-$3)/SO(D$-$2), which is the "R-symmetry" gauged in conformal supergravity.

## 3. SUPERSYMMETRY

In the bosonic case, $e_A{}^M$ is a (2D+n)×(2D+n) matrix, representing graviton, axion, and n types of photons (with dilaton in $\Phi$) in D dimensions. The left- and right-handed indices are $\mathcal{A} = a$ and $\tilde{\mathcal{A}} = (\tilde{a}, \hat{a})$, where $a$ and $\tilde{a}$ take D values and $\hat{a}$ takes n. The flat-space values of the fields are

$$\langle e_A{}^M \rangle = \begin{pmatrix} \delta_a^m & \tfrac{1}{2}\eta_{am} & 0 \\ \delta_{\tilde{a}}^m & -\tfrac{1}{2}\eta_{\tilde{a}m} & 0 \\ 0 & 0 & \delta_{\hat{a}}^{\hat{m}} \end{pmatrix} \quad \Rightarrow \quad \langle g_{AB} \rangle \equiv \eta_{AB} = \begin{pmatrix} \eta_{ab} & 0 & 0 \\ 0 & -\eta_{ab} & 0 \\ 0 & 0 & -\delta_{\hat{a}\hat{b}} \end{pmatrix}$$

We next discuss the supersymmetric case. Since we consider only the classical mechanics of superstrings, we can treat D=3,4,6,10 dimensions for the supersymmetric modes of heterotic superstrings. (The generalization to type II superstrings will not be considered here.) Therefore, we consider background fields describing N=1 supergravity (+ a tensor multiplet, in the cases where pure supergravity doesn't include an axion) together with (abelian) Yang-Mills multiplets.

We now have $\mathcal{A} = (\alpha, a, \tilde{\alpha})$ (and still $\tilde{\mathcal{A}} = (\tilde{a}, \hat{a})$), with $\Pi_{\mathcal{A}} = (\Pi_\alpha, \Pi_a, \Pi_{\tilde{\alpha}})$ the curved superspace generalization of the affine Lie algebra $(D_\alpha, P_a, \Omega^\alpha)$ introduced in [3]. ($\Omega^\alpha$ is necessary to make the range of the $A$ and $M$ indices agree, to define an invertible $g_{\mathcal{AB}}$, and to express $L_+$ as $g\Pi\Pi$.) The generalization of the bosonic tangent-space vacuum metric is then

$$\langle g_{AB} \rangle \equiv \eta_{AB} = \begin{pmatrix} 0 & 0 & \delta_\alpha^\beta & 0 & 0 \\ 0 & \eta_{ab} & 0 & 0 & 0 \\ -\delta_\beta^\alpha & 0 & 0 & 0 & 0 \\ 0 & 0 & 0 & -\eta_{ab} & 0 \\ 0 & 0 & 0 & 0 & -\delta_{\hat{a}\hat{b}} \end{pmatrix}$$

We have the usual superspace coordinates $x^{\mathbf{M}} = (\theta^\mu, x^m)$.

In addition to the Virasoro operators, which are first-class constraints, we now also have the second-class constraint $\Pi_\alpha$. As a result, the would-be GL(D|2D')⊗GL(D+n) gauge symmetry is restricted so that $\Pi_\alpha$ transforms into itself. Furthermore, we will see in the following section that we need to further restrict this tangent-space symmetry $\lambda_\alpha{}^\beta$ on $\Pi_\alpha$ from GL(D') to the largest symmetry that leaves $\gamma$-matrices invariant. In D=3,4,6 this group is GL(2,A) for A = real, complex, quaternion; and we have GL(2,A) = Lorentz⊗scale⊗internal, with internal = SO(D−2)/SO(D−3) = $S_{D-3}$ = I,U(1),SU(2). (Perhaps there is some way to generalize to octonion for D=10. Otherwise, it seems that there is no internal symmetry in that case.) Thus, in D=3 there is no restriction on $\lambda_\alpha{}^\beta$; in D=4, using 4D Weyl spinor notation to write the spinor index as $(\alpha, \dot{\alpha})$, we have $\lambda_\alpha{}^{\dot\beta} = \lambda_{\dot\alpha}{}^\beta = 0$; in D=6, using SU(2) Majorana-Weyl spinor notation to write the spinor index as $i\alpha$, we have $\lambda_{i\alpha}{}^{j\beta} = \delta_i{}^j \lambda_\alpha{}^\beta + \delta_\alpha{}^\beta \lambda_i{}^j$; and in D=10 we can write $\lambda_\alpha{}^\beta = \delta_\alpha{}^\beta \lambda + \frac{1}{2} \gamma^{\check{a}\check{b}}{}_\alpha{}^\beta \lambda_{\check{a}\check{b}}$, where $\check{a}$ is a Lorentz vector index. We will also find that $g_{\alpha\beta}$ is restricted so $\{\Pi_\alpha, \Pi_\beta\}$ doesn't generate $\delta'$ terms. The net result of all these restrictions is:

$$g_{\alpha\beta} = 0, \quad \lambda_\alpha{}^b = \lambda_\alpha{}^{\tilde\beta} = \Pi_\alpha{}^\beta{}_\delta{}^\gamma \lambda_\gamma{}^\delta = 0$$

where $\Pi_\alpha{}^\beta{}_\delta{}^\gamma$ is a projection operator that picks out the pieces of $\lambda_\alpha{}^\beta$ we want to discard, as just described. The full GL(D|2D')⊗GL(D+n) gauge invariance could be restored by defining the second-class constraint to be $A_\alpha = A_\alpha{}^B \Pi_B$ for some $A_\alpha{}^B$, and introducing a new symmetry group (Lorentz, internal, scale) acting on $A_\alpha$. The previous group would then be obtained in the gauge $A_\alpha{}^B = \delta_\alpha{}^B$.

As in ordinary superspace, the restriction of the tangent-space group from the naive generalization of ordinary gravity allows the existence of nontrivial torsions (while reducing the set of curvatures). Unlike ordinary superspace, we will find that *all* such torsions must either be constrained to vanish or be found to be pure gauge. Since the $F$'s are already $\lambda^M$-covariant, we start with them and consider $\lambda_A{}^B$ covariance. (Equivalently, we can look at the torsions defined below and analyze those that are connection-independent due to these new restrictions.) Examining the $F$ transformation law from the previous section, we see that a specific component of $F_{AB}{}^C$ will be covariant if, as a consequence of our conditions on $\lambda_A{}^B$ and $g_{AB}$ (including those from the previous section),

$$\lambda_A{}^C = \lambda_B{}^C = \lambda_A{}^D g_{DB} = 0$$

for that specific choice of $A, B, C$. The last constraint is $C$-independent, and implies $AB = \mathcal{A}\tilde{\mathcal{B}}$ or $\alpha\beta$. (These are exactly the values for which $g_{AB} = 0$.) The remaining constraints then imply $\mathcal{A} = \alpha$ and $C = c$ or $\tilde{\gamma}$ in the former case, and $C =$ anything but $\gamma$ in the latter case, but both cases allow $C = \gamma$ when the spinor indices are appropriately projected. The final result is then that the fully covariant $F$'s are

$$F_{\alpha\beta}{}^{\tilde{\mathcal{C}}}, \ F_{\alpha\beta}{}^{c,\tilde{\gamma}}, \ \Pi_{\alpha\beta}{}^{\gamma}{}_{\delta}{}^{\epsilon\zeta} F_{\epsilon\zeta}{}^{\delta}, \ F_{\alpha\tilde{\mathcal{B}}}{}^{c,\tilde{\gamma}}, \ \Pi_{\alpha}{}^{\gamma}{}_{\delta}{}^{\epsilon} F_{\epsilon\tilde{\mathcal{B}}}{}^{\delta}$$

where we have written them in groups which transform into themselves under the tangent-space transformations. $\Pi_{\alpha\beta}{}^{\gamma}{}_{\delta}{}^{\epsilon\zeta}$ is another projection operator which performs a similar function; for example, in D=4 it picks out $F_{\alpha\beta}{}^{\dot{\gamma}}$ and $F_{\dot{\alpha}\dot{\beta}}{}^{\gamma}$.

## 4. CONSTRAINTS

The introduction of massless background fields into the Green-Schwarz superstring for the purpose of deriving constraints on the background was first performed by Witten [4], who worked in the lagrangian formalism, effectively calculating the closure of the algebra of the first-class constraints $L_\pm$ (the Virasoro operators) and $B^\alpha$ (the generator of $\kappa$ symmetry). Shapiro and Taylor [5] worked in the hamiltonian formalism, and calculated also the covariance of the second-class constraint. Our calculation will be similar to theirs, but in duality covariant form.

We first consider the Virasoro algebras. Checking $[L_\pm, L_\pm] \approx 0$ (where "$\approx$" means modulo generators of the constraint algebra $L_\pm$ and $\Pi_\alpha$) is easy: They give no constraints on the background, but only verify the expressions for $L_\pm$ in terms of $g^{\mathcal{AB}}$ and $g^{\tilde{\mathcal{A}}\tilde{\mathcal{B}}}$. $[L_+, L_-] \approx 0$ is then trivial: Just use $[L_\pm, L_\pm]$ and $L_+ + L_- = \frac{1}{2} Z^M Z_M$.

Witten effectively calculated (in addition to the Virasoro algebras) $[L_\pm, B^\alpha] \approx 0$ and $\{B^\alpha, B^\beta\} \approx 0$. Shapiro and Taylor calculated the slightly stronger constraints $[L_\pm, \Pi_\alpha] \approx 0$ and $\{B^\alpha, \Pi_\beta\} \approx 0$. We will consider conditions that are slightly stronger yet, which in the conventional gauge are $[L_\pm, \Pi_\alpha] \approx 0$ and $\{\Pi_\alpha, \Pi_\beta\} \approx i\delta \gamma^a_{\alpha\beta} \Pi_a$. The calculation of $[L_\pm, B^\alpha]$ and $[L_\pm, \Pi_\alpha]$ uses just $[\Pi_{\tilde{\mathcal{A}}}, \Pi_\alpha]$, while $\{B^\alpha, B^\beta\}$ and $\{B^\alpha, \Pi_\beta\}$ use just $\{\Pi_\alpha, \Pi_\beta\}$.

Invariance of the second-class constraints under conformal transformations requires
$$[L_-, \Pi_\alpha] \approx 0 \quad \Rightarrow \quad F_{\alpha\tilde{\mathcal{B}}}{}^{c,\tilde{\gamma}} = 0$$
$[L_+, \Pi_\alpha] \approx 0$ can then be checked easily using $L_+ + L_- = \frac{1}{2} Z^M Z_M$. We also require that the second-class constraints satisfy the same type of algebra as in empty space: For some vector operator $\widehat{\Pi}_a$,
$$\{\Pi_\alpha, \Pi_\beta\} \approx i\delta F_{\alpha\beta}{}^c \widehat{\Pi}_c$$
This can be obtained by constraining
$$g_{\alpha\beta} = F_{\alpha\beta}{}^{\tilde{c}} = F_{\alpha\beta\gamma} = 0$$
$$\Rightarrow \quad F_{\alpha\beta}{}^{\tilde{\gamma}} = -F_{\alpha\beta}{}^c g_{c\delta}(g_{\tilde{\gamma}\delta})^{-1} = F_{\alpha\beta}{}^c (g^{cd})^{-1} g^{d\tilde{\gamma}}, \quad \widehat{\Pi}_a = \Pi_a + (g^{ab})^{-1} g^{b\tilde{\alpha}} \Pi_{\tilde{\alpha}}$$
Thus, of all the tensors that are first-order in derivatives, only $F_{\alpha\beta}{}^c$ remains undetermined. The last result can also be written as
$$\{\Pi_\alpha, \Pi_\beta\} \approx i\delta F_{\alpha\beta}{}^c (g^{cd})^{-1} \Pi^c$$

As a consequence of these constraints, the Bianchi identities imply
$$0 = F_{(\alpha\beta}{}^E F_{\gamma\delta)E} = F_{(\alpha\beta}{}^e (g^{ef})^{-1} F_{\gamma\delta)}{}^f$$
which is the generalization of the $\gamma$-matrix identity $\gamma_{a(\alpha\beta} \gamma^a{}_{\gamma\delta)} = 0$ for D=3,4,6,10.

In 3D (D'=2) the symmetric $\alpha\beta$ indices on $F_{\alpha\beta}{}^c$ represent a vector index under GL(D')=SO(2,1)⊗GL(1) (Lorentz and scale transformations). Thus there is a GL(D) gauge in which $F_{\alpha\beta}{}^c$ becomes the usual 3D $\gamma$-matrices. For a similar procedure to work in general, we need to further restrict the tangent-space symmetry $\lambda_\alpha{}^\beta$ on $\Pi_\alpha$ from GL(D') to the largest symmetry that leaves the $\gamma$-matrices invariant, as discussed in the previous section. Then $F_{\alpha\beta}{}^c$ can be separated into a $\gamma$-matrix piece and the rest, which is constrained to vanish to "preserve representations." The additional constraint is then
$$F_{\alpha\beta}{}^{c,\tilde{\gamma}} = \gamma_{\alpha\beta}^{\check{d}} F_{\check{d}}{}^{c,\tilde{\gamma}}$$
for some $F_{\check{d}}{}^{c,\tilde{\gamma}}$, where $\gamma_{\alpha\beta}^{\check{d}}$ are the usual $\gamma$-($\sigma$-)matrices, invariant under GL(2,A). Using the Bianchi identity just derived, this implies
$$\eta^{\check{a}\check{b}} F_{\check{a}}{}^c F_{\check{b}}{}^d \sim g^{cd}$$
which states that, up to a scale factor, $F_{\check{a}}{}^b$ is a vierbein for $g^{ab}$.

In summary, the constraints are
$$g_{\alpha\beta} = F_{\alpha\beta}{}^{\tilde{\mathcal{C}}} = F_{\alpha\beta\gamma} = F_{\alpha\tilde{\mathcal{B}}}{}^{c,\tilde{\gamma}} = 0, \quad F_{\alpha\beta}{}^{c,\tilde{\gamma}} = \gamma_{\alpha\beta}^{\check{d}} F_{\check{d}}{}^{c,\tilde{\gamma}}$$
It then follows from the Bianchi identities that
$$\Pi_{\alpha\beta\gamma}{}^{\epsilon\zeta}_\delta F_{\epsilon\zeta}{}^\delta = \Pi_\alpha{}^\gamma{}_\delta{}^\epsilon F_{\epsilon\tilde{\mathcal{B}}}{}^\delta = 0$$
Thus, all torsions are determined except $F_{\check{a}}{}^b$, which can be fixed to $\delta_a^b$ by a GL(D) transformation $\lambda_a{}^b$.

We can now define the $\kappa$-symmetry generators $B^\alpha$ in terms of a new tensor $h^{\mathcal{ABC}}$ ($h^{(\mathcal{AB})\mathcal{C}} = 0$), with only $h^{\alpha\beta\mathcal{C}}$ nonvanishing (a covariant condition), as
$$h^{\alpha\beta}{}_\gamma = 0 \quad \Rightarrow \quad h^{\alpha\beta\tilde{\gamma}} = h^{\alpha\beta c}(g^{cd})^{-1} g^{d\tilde{\gamma}}$$
$$\Rightarrow \quad B^\alpha = h^{\alpha\beta\mathcal{C}} \Pi_\mathcal{C} \Pi_\beta = h^{\alpha\beta c} \widehat{\Pi}_c \Pi_\beta + h^{\alpha\beta\gamma} \Pi_\gamma \Pi_\beta$$
$$\Rightarrow \quad \{B^\alpha, \Pi_\beta\} \approx i\delta h^{\alpha\gamma c} F_{\beta\gamma}{}^d \widehat{\Pi}_d \widehat{\Pi}_c$$
Using the identity
$$\tfrac{1}{2} g^{ab} \widehat{\Pi}_b \widehat{\Pi}_a = \tfrac{1}{2} g^{ab} \Pi_b \Pi_a + g^{a\tilde{\beta}} \Pi_{\tilde{\beta}} \Pi_a + \tfrac{1}{2} g^{\tilde{\alpha}\tilde{\beta}} \Pi_{\tilde{\beta}} \Pi_{\tilde{\alpha}} \approx \tfrac{1}{2} g^{\mathcal{AB}} \Pi_\mathcal{B} \Pi_\mathcal{A} = L_+ \approx 0$$
we then find the condition on $h$
$$h^{\alpha\gamma(c} F_{\beta\gamma}{}^{d)} \sim g^{cd}$$
which is the generalization of the $\gamma$-matrix identity $\gamma^{(c\alpha\gamma} \gamma_{\beta\gamma}^{d)} = 2\eta^{cd} \delta^\alpha_\beta$. The solution to this condition is
$$h^{\alpha\beta c} = \gamma^{\check{d}\alpha\beta} F_{\check{d}}{}^c$$
In fact, even without the earlier representation-preserving constraint on $F_{\alpha\beta}{}^{c,\tilde{\gamma}}$, the condition implies that $F_{\alpha\beta}{}^c$ and $h^{\alpha\beta c}$ together form the Dirac matrices $\Gamma = \begin{pmatrix} 0 & h \\ \gamma & 0 \end{pmatrix}$ up to a unitary transformation [5]; however, at least for D=4 and 6, an explicit expression for that unitary matrix would be required, to separate chiral and antichiral spinors so that chirality constraints on $\Phi$ can be imposed. Therefore, we use the stronger representation-preserving constraint on $F_{\alpha\beta}{}^c$ give above.

## 5. CONVENTIONAL GAUGE

We first consider the bosonic case. There are two ways to choose duality-preserving gauges for the tangent-space GL(D)⊗GL(D+n) gauge invariance: (1) Choose the gauge $g_{AB} = \eta_{AB}$, leaving the coset space SO(D,D+n)/SO(D−1,1)⊗SO(D+n−1,1), with residual local invariance SO(D−1,1)⊗SO(D+n−1,1), in the style of Duff [6]. Explicit solution of the gauge condition $g_{AB} = \eta_{AB}$ requires breaking manifest duality covariance. (2) Solve $g_{a\tilde{\mathcal{B}}} = 0$ for $e_{\tilde{\mathcal{A}}m}$, and use GL(D+n) to fix the rest of $e_{\tilde{\mathcal{A}}M}$, which leaves for tangent-space invariance just GL(D), in the style of Maharana and Schwarz [7]. In this gauge one works with just $\Pi_a$, which is unconstrained. (Duff and Maharana and Schwarz considered these reduced gauge groups for the d dimensions where the fields were constant.)

The usual vierbein formalism can be obtained from either of these gauges by further gauge fixing with only SO(D−1,1) gauge invariance remaining in the tangent space, in a way which breaks manifest duality invariance but preserves the component $\lambda^M$ gauge invariances: We compare the transformation of $e_A{}^M$

$$\delta e_A{}^m = (\lambda^n \partial_n e_A{}^m - e_A{}^n \partial_n \lambda^m)$$
$$\delta e_{Am} = (\lambda^n \partial_n e_{Am} + e_{An} \partial_m \lambda^n) + e_A{}^n \partial_{[m} \lambda_{n]} - e_{A\hat{n}} \partial_m \lambda^{\hat{n}}$$
$$\delta e_A{}^{\hat{m}} = (\lambda^n \partial_n e_A{}^{\hat{m}}) - e_A{}^n \partial_n \lambda^{\hat{m}}$$

with the usual gauge transformations

$$\delta e_a{}^m = (\lambda^n \partial_n e_a{}^m - e_a{}^n \partial_n \lambda^m)$$
$$\delta b_{mn} = (\lambda^p \partial_p b_{mn} - b_{p[m} \partial_{n]} \lambda^p) - \partial_{[m} \lambda_{n]} + A_{[m\hat{p}} \partial_{n]} \lambda^{\hat{p}}$$
$$\delta A_m{}^{\hat{n}} = (\lambda^p \partial_p A_m{}^{\hat{n}}) - \partial_m \lambda^{\hat{n}}$$

to identify

$$\Pi_a = e_a{}^m [P_m + A_m{}^{\hat{b}} Z_{\hat{b}} + X'^n \tfrac{1}{2}(b_{nm} + A_{n\hat{b}} A_{m\hat{b}} + g_{nm})]$$
$$\Pi_{\tilde{a}} = e_a{}^m [P_m + A_m{}^{\hat{b}} Z_{\hat{b}} + X'^n \tfrac{1}{2}(b_{nm} + A_{n\hat{b}} A_{m\hat{b}} - g_{nm})]$$
$$\Pi_{\hat{a}} = Z_{\hat{a}} - X'^m A_{m\hat{a}}$$

with $g^{ab} = -g^{\tilde{a}\tilde{b}} = \eta^{ab}$ and $g^{\hat{a}\hat{b}} = -\delta_{\hat{a}\hat{b}}$ in $L_{\pm}$.

The supersymmetric case is treated similarly: (1) Use the $g_{AB}$ constraints and some of the $\lambda_A{}^B$ gauge invariance to fix $g_{AB} = \eta_{AB}$; (2) use most of the remaining $\lambda_A{}^B$ gauge invariance to eliminate fields which are invariant under $\lambda^M$ transformations (or to fix fields which transform in the same way to be equal). The result is:

$$\Pi_\alpha = \Upsilon_\alpha$$
$$\Pi_a = \Upsilon_a + \tfrac{1}{2} X'^{\mathbf{M}} e_{\mathbf{M}}{}^b \eta_{ba}$$
$$\Pi_{\tilde{a}} = \Upsilon_a - \tfrac{1}{2} X'^{\mathbf{M}} e_{\mathbf{M}}{}^b \eta_{ba}$$
$$\Pi_{\hat{a}} = Z_{\hat{a}} - X'^{\mathbf{M}} A_{\mathbf{M}\hat{a}} - W^\alpha{}_{\hat{a}} \Pi_\alpha$$
$$\Pi_{\tilde{\alpha}} = X'^{\mathbf{M}} e_{\mathbf{M}}{}^\alpha + W^{\alpha \hat{a}} \Pi_{\hat{a}} - \tfrac{1}{2} W^\alpha{}_{\hat{a}} W^\beta{}_{\hat{a}} \Pi_\beta$$

$$\Upsilon_{\mathbf{A}} \equiv e_{\mathbf{A}}{}^{\mathbf{M}} [ P_{\mathbf{M}} + A_{\mathbf{M}}{}^{\hat{b}} Z_{\hat{b}} + X'^{\mathbf{N}} \tfrac{1}{2}(b_{\mathbf{NM}} + A_{\mathbf{M}\hat{b}} A_{\mathbf{N}\hat{b}})]$$

where $e_{\mathbf{A}}{}^{\mathbf{M}}$ is the usual superspace vielbein, and $e_{\mathbf{M}}{}^{\mathbf{A}}$ its inverse. We also have the super 2-form $b_{\mathbf{MN}}$, the Yang-Mills supervector $A_{\mathbf{M}}{}^{\hat{a}}$, and a spinor $W^{\alpha\hat{a}}$ which will be determined to be the Yang-Mills field strength by the torsion constraints, and which appears in the affine Lie algebra as the $\tilde{\alpha}$ component of $A_{\mathcal{A}}{}^{\hat{a}}$ [3]. (There is also the gauge where $W$ appears in $\Pi$ only as the linear term in $\Pi_{\tilde{\alpha}}$ if we allow $g_{\tilde{\alpha}\hat{a}} = W^\alpha{}_{\hat{a}}$ and $g_{\tilde{\alpha}\tilde{\beta}} = -W^\alpha{}_{\hat{a}} W^\beta{}_{\hat{a}}$.) This gauge is not manifestly duality covariant because it treats $e_A{}^M$ and $e_{AM}$ differently. The remaining $\lambda_A{}^B$ gauge invariance is given by just $\lambda_\alpha{}^\beta$ and $\lambda_{[ab]} = \lambda_{[\tilde{a}\tilde{b}]}$.

In this gauge, the duality covariants reduce to

$$F_{\alpha\beta}{}^{\tilde{\gamma}} = \tfrac{1}{2} H_{\alpha\beta\gamma}$$
$$F_{\alpha\beta}{}^c = \tfrac{1}{2}(c_{\alpha\beta}{}^c - H_{\alpha\beta}{}^c)$$
$$F_{\alpha\beta}{}^{\tilde{c}} = F_\alpha{}^{\tilde{c}\tilde{\beta}} = \tfrac{1}{2}(c_{\alpha\beta}{}^c + H_{\alpha\beta}{}^c)$$
$$F_{\alpha\beta}{}^{\hat{c}} = -F_{\alpha\hat{c}}{}^{\tilde{\beta}} = \mathcal{F}_{\alpha\beta}{}^{\hat{c}}$$
$$F_{\alpha\tilde{b}}{}^c = \tfrac{1}{2}(c_{\alpha(b}{}^{c)} - H_{\alpha b}{}^c)$$
$$F_{\alpha\hat{b}}{}^c = -\mathcal{F}_\alpha{}^c{}_{\hat{b}} + F_{\alpha\delta}{}^c W^\delta{}_{\hat{b}}$$

where we have defined the usual duality-noncovariant field strengths

$$c_{\mathbf{AB}}{}^{\mathbf{C}} \equiv (e_{[\mathbf{A}} e_{\mathbf{B})}{}^{\mathbf{M}}) e_{\mathbf{M}}{}^{\mathbf{C}}$$
$$H_{\mathbf{ABC}} \equiv e_{\mathbf{A}}{}^{\mathbf{M}} e_{\mathbf{B}}{}^{\mathbf{N}} e_{\mathbf{C}}{}^{\mathbf{P}} (\tfrac{1}{2} \partial_{[\mathbf{M}} b_{\mathbf{NP})} + A_{[\mathbf{M}\hat{d}} \partial_{\mathbf{N}} A_{\mathbf{P})\hat{d}})$$
$$\mathcal{F}_{\mathbf{AB}}{}^{\hat{c}} \equiv e_{\mathbf{A}}{}^{\mathbf{M}} e_{\mathbf{B}}{}^{\mathbf{N}} \partial_{[\mathbf{M}} A_{\mathbf{N})}{}^{\hat{c}}$$

In the conventional gauge, since $g^{ab} = \eta^{ab}$, we can use the $\lambda_{[ab]}$ invariance to fix $F_{\hat{a}}{}^b = \psi \delta_a^b$. The remaining tangent-space gauge invariance is just (the constrained) $\lambda_\alpha{}^\beta$. Finally, we can use the scale part $\lambda_\alpha{}^\alpha$ to fix $\psi = 1$. This leaves local Lorentz and internal symmetries.

The conventional-gauge form of the constraints is then

$$H_{\alpha\beta\gamma} = 0$$

$$\mathcal{F}_{\alpha\beta\hat{c}} = 0, \quad c_{\alpha\beta}{}^c = -H_{\alpha\beta}{}^c = \gamma^c_{\alpha\beta}$$

$$H_{\alpha bc} = c_{\alpha(bc)} = 0, \quad \mathcal{F}_{\alpha b}{}^{\hat{c}} = \gamma_{b\alpha\beta} W^{\beta\hat{c}}$$

This is the form of the constraints that comes naturally from the superstring [8]. As in the bosonic case, making a field redefinition (superscale transformation) of the vielbein to make the fields appear in more conventional form would also complicate the form of the duality transformations.

## 6. PREPOTENTIALS

In D=3 and 4, these are the off-shell constraints for N=1 conformal supergravity coupled to a physical tensor multiplet, in the superscale gauge where the tensor multiplet is gauged to a constant. This is the "string gauge" [9], and also can be related directly to the usual Weyl superscale gauge for supergravity by field redefinitions equivalent to a change of superscale gauge. In the bosonic case, this field redefinition is the usual one which strips the dilaton dependence from the curvature term in the action. The compensator multiplet, which contains the degrees of freedom which complete conformal supergravity to ordinary supergravity, does not appear in $e_A{}^M$, but appears separately through coupling to the world-sheet curvature or reparametrization ghosts, just as in the bosonic case. This coupling has not been derived in the usual Green-Schwarz formalism, since coupling to the world-sheet curvature (and probably also to the Lagrange multiplier for $\kappa$ symmetry) cannot be shown to be gauge invariant without considering one-loop effects, which are not fully understood in the covariant Green-Schwarz formalism; and the covariant ghost structure of that formalism, and therefore coupling to those ghosts, is also not completely understood. However, the curvature coupling in D=4 has been described in a modified Green-Schwarz formalism by Berkovits [10], and the result agrees with that obtained by other arguments [9]: The compensator multiplet is a chiral scalar superfield, which gives the auxiliary field structure of "old minimal" supergravity. In D=6, the (self-dual) tensor multiplet is itself a compensator multiplet, though it appears with the correct sign in the action to describe physical fields [11]. (In D=10 the situation is not clear, since the superconformal group doesn't exist there.)

In D=3 the duality covariant torsion constraints of section 4 can be solved in a manifestly duality covariant way in terms of the components $e_\alpha{}^M$ of the vielbein, where $e_\alpha{}^{\mathbf{M}}$ is the unconstrained superfield describing 3D supergravity, as usual [12] (the analog to $e_a{}^m$ in the bosonic case), $e_{\alpha\mathbf{M}}$ describes the tensor multiplet (the analog to $e_{am}$), and $e_\alpha{}^{\hat{m}}$ describes the vector multiplets, as usual (the analog to $e_a{}^{\hat{m}}$). In D=4 there is also a representation-preserving constraint to solve: Again using 4D Weyl spinor notation to write the spinor index as $(\alpha, \dot{\alpha})$, we solve this constraint as usual [13]:

$$\{\Pi_\alpha, \Pi_\beta\} = i\delta F_{\alpha\beta}{}^\gamma \Pi_\gamma, \quad \{\Pi_{\dot\alpha}, \Pi_{\dot\beta}\} = i\delta F_{\dot\alpha\dot\beta}{}^{\dot\gamma} \Pi_{\dot\gamma}$$

$$\Rightarrow \quad \Pi_\alpha = A_\alpha{}^\mu e^W Z_\mu e^{-W}, \quad \Pi_{\dot\alpha} = A_{\dot\alpha}{}^{\dot\mu} e^{-\overline{W}} Z_{\dot\mu} e^{\overline{W}}, \quad W = \int W^M Z_M$$

We then have the analogous expression for the derivations

$$e_\alpha = A_\alpha{}^\mu e^w \partial_\mu e^{-w}, \quad w = W^M i\partial_M$$

(We have used the fact that $F_{\alpha\beta}{}^{\dot\gamma} = 0$ follows from the other constraints [14] by the $F$ Bianchi identities. The expression for $e_\alpha$ determines $e_\alpha{}^M$ only up to $\partial^M f_\alpha$ terms, but $e_\alpha{}^M$ is unambiguously determined by the expression for $\Pi_\alpha$.) $W^{\mathbf{M}}$ is the usual unconstrained superfield (prepotential) of 4D N=1 supergravity, $W_{\mathbf{M}}$ describes the tensor multiplet [10], and $W^{\hat{m}}$ is the usual prepotential for vector multiplets. Duality then acts on the $M$ index of $W^M$ in the obvious way. $W$ is basically a complexification of the group element $\Lambda$. It has the (finite) gauge transformation

$$e^{W'} = e^{-i\Lambda} e^W e^{i\Xi}, \quad \Xi = \int \xi^M Z_M, \quad \partial_\mu \xi^M = 0 \; except \; \partial_\mu \xi^\nu \neq 0$$

where the new (mostly) antichiral parameter $\Xi$ gives an invariance because its exponentials can push past the $Z_\mu$ in $\Pi_\mu$ (or $\partial_\mu$ in $e_\mu$) to cancel, and $\xi^\mu$ is unconstrained because $A_\alpha{}^\mu$ transforms to cancel that piece. There is also the additional gauge transformation

$$W'^M = W^M + \partial^M \zeta$$

which leaves $W$ invariant (as the gauge transformation for gauge transformation did for $\Lambda$). As in supergravity, it is generally convenient to work with the $\Lambda$-invariant combination

$$e^U \equiv e^{\overline{W}} e^W$$

in terms of which $\Pi$ can be expressed by a suitable nonunitary similarity transformation. ($U$ is basically the real part of $W$.)

Berkovits [10] used the Ogievetsky-Sokatchev form of $U^m$ [15], which is related nonlinearly to ours [16], and thus the resulting duality transformations for his formalism [17] are also nonlinear in it (and so nonmanifest). His choice of fields corresponds to choosing the gauge $W^\mu = W^{\dot\mu} = W_\mu = 0$ (using the gauge parameters $\lambda^\mu$, $\xi^\mu$, and $\xi_\mu$). In this gauge the constraints $g_{\alpha\beta} = g_{\dot\alpha\dot\beta} = 0$ are satisifed automatically, while he imposed a constraint equivalent to $g_{\alpha\dot\beta} = 0$.

In general dimensions the constraint $g_{\alpha\beta} = 0$ (switching back to arbitrary-D spinor notation) can be solved explicitly by imposing an appropriate antisymmetry condition on $e_{\alpha\mu}$. Although this solution does not break manifest duality, since it does not effect the vector indices on which SO(D,D+n) acts, we can instead treat the constraint as a field equation. In fact, in D=3 all the torsion constraints can also be taken as field equations, although there is no advantage, since they all are "conventional" constraints which simply determine redundant variables in terms of $e_\alpha{}^M$. Similar remarks apply in D=4, except for the representation-preserving constraint, which is necessary to write chiral actions with $\Phi$, and introduces $U^M$ as the basic field. In D=4 we already treat the $\mu$ indices differently when solving the representation-preserving constraint in terms of $W$. (In principle we could also treat $g_{\mathcal{A}\tilde{\mathcal{B}}} = 0$ as a field equation, even in the bosonic case.)

The four-dimensional case of this superspace will be considered in more detail in a future paper.

## 7. COVARIANT DERIVATIVES

We now want to construct covariant derivatives in analogy to ordinary gravity,

$$\nabla_A = e_A + \omega_{A\mathcal{B}}{}^{\mathcal{C}} G_{\mathcal{C}}{}^{\mathcal{B}} + \omega_{A\tilde{\mathcal{B}}}{}^{\tilde{\mathcal{C}}} G_{\tilde{\mathcal{C}}}{}^{\tilde{\mathcal{B}}}$$

where $G_{\mathcal{C}}{}^{\mathcal{B}}$ and $G_{\tilde{\mathcal{C}}}{}^{\tilde{\mathcal{B}}}$ are the generators of the left- and right-handed GL invariances. They act on all tangent-space indices in the usual way:

$$G_{\mathcal{A}}{}^{\mathcal{B}} V_{\mathcal{C}} = \delta_{\mathcal{C}}{}^{\mathcal{B}} V_{\mathcal{A}}, \quad G_{\mathcal{A}}{}^{\mathcal{B}} V^{\mathcal{C}} = -\delta_{\mathcal{A}}{}^{\mathcal{C}} V^{\mathcal{B}}$$

(In the supersymmetric case $\omega_{A\mathcal{B}}{}^{\mathcal{C}}$ is restricted on its group indices, so the only parts of $G_{\mathcal{A}}{}^{\mathcal{B}}$ that contribute to the covariant derivative are the ones appropriate to the restricted group.) Since the commutator of two $\Pi$'s includes $\delta'$ terms, a more convenient way to define torsions and curvatures is through the analysis of group elements with tangent-space indices $\lambda^A$. We first convert the expression for the Lie derivative to tangent-space indices by converting coordinate-space indices with the vielbein ($\lambda^M = \lambda^A e_A{}^M$), and adding and subtracting GL-connection terms to convert derivatives into covariant ones:

$$\lambda^A_{[1,2]} = \lambda^B_{[1} \nabla_B \lambda^A_{2]} - \tfrac{1}{2} \lambda_{[1B} \nabla^A \lambda^B_{2]} + \lambda^B_1 \lambda^C_2 T_{BC}{}^A$$

$$\Rightarrow \quad T_{AB}{}^C = F_{AB}{}^C + (\omega_{[AB)}{}^C + \tfrac{1}{2}\omega^C{}_{[AB)})$$
$$= K_{[AB)}{}^C + \tfrac{1}{2}K^C{}_{[AB)}, \quad K_{AB}{}^C = f_{AB}{}^C + \omega_{AB}{}^C$$

This torsion can be set consistently to zero in the bosonic case. In the supersymmetric case, where the tangent-space group is restricted to be smaller than these GL groups, some torsions do not contain connections; these torsions were constrained in section 4. Furthermore, we can set the tangent-space metric to be covariantly constant:
$$\omega\text{-dependent } T_{AB}{}^C = \nabla_A g_{BC} = 0$$
$$\Rightarrow \quad \omega_{\mathcal{A}\tilde{\mathcal{B}}}{}^{\tilde{\mathcal{C}}} = -F_{\mathcal{A}\tilde{\mathcal{B}}}{}^{\tilde{\mathcal{C}}}, \quad \omega_{\mathcal{A}(\mathcal{B}\mathcal{C})} = -e_{\mathcal{A}}g_{\mathcal{B}\mathcal{C}}, \quad \omega_{[\mathcal{A}\mathcal{B}\mathcal{C})} = -\tfrac{1}{3}F_{[\mathcal{A}\mathcal{B}\mathcal{C})}$$
and similarly for $\mathcal{A} \leftrightarrow \tilde{\mathcal{A}}$. (The two constraints are partly redundant, since $T_{AB}{}^C = 0 \Rightarrow \nabla_{[A}g_{B)C} = 0$.) Unfortunately, there are no corresponding $F$'s for the remaining connection $\omega_{\mathcal{A}[\mathcal{B}\mathcal{C})} + \omega_{\mathcal{B}[\mathcal{A}\mathcal{C})}$, since $\omega_{\mathcal{A}\mathcal{B}\mathcal{C}}$ (and $\delta\omega_{\mathcal{A}\mathcal{B}\mathcal{C}}$) has the full tensor structure of an arbitrary third-rank tensor, while $F_{\mathcal{A}\mathcal{B}\mathcal{C}}$ contains only $F_{[\mathcal{A}\mathcal{B}\mathcal{C})}$ and $e_{\mathcal{A}}g_{\mathcal{B}\mathcal{C}}$.

However, because we have an integration measure $\Phi^2$, we can define $\nabla$ to satisfy integration by parts
$$\int \Phi^2 V^A \nabla_A A = -\int \Phi^2 A \nabla_A V^A \quad \Leftrightarrow \quad \tilde{T}_A \equiv \Phi^2 \overleftarrow{\nabla}_A \Phi^{-2} = 0$$
$$\Leftrightarrow \quad \omega_{BA}{}^B = -\tilde{F}_A \equiv -\Phi^2 \overleftarrow{e}_A \Phi^{-2}$$
for arbitrary $A$ and $V^A$. This new $F$ satisfies the Bianchi identities
$$e_{[A}\tilde{F}_{B)} - e_C F_{AB}{}^C - \tilde{F}_C F_{AB}{}^C = 0$$
$$e_A \tilde{F}^A + \tfrac{1}{2}\tilde{F}_A{}^2 + \tfrac{1}{2}e_A e_B g^{AB} - \tfrac{1}{432}F_{[ABC)}{}^2 - \tfrac{1}{8}(e^A g^{BC})(e_B g_{AC}) = 0$$
Another reason this new torsion $\tilde{T}_A$ occurs is because we have $\partial^M \partial_M = 0$ as well as $\partial_{[M}\partial_{N)} = 0$. Thus, besides the usual torsion, which is generated by commutators, we have a new torsion generated by a d'Alembertian:
$$0 = \partial_M \partial^M A = (e^A A)\overleftarrow{e}_A = (\nabla^A A)\overleftarrow{\nabla}_A \quad \Rightarrow \quad \nabla_A \nabla^A A = -\tilde{T}_A \nabla^A A$$

This is the analog of the usual relation
$$[\nabla_A, \nabla_B\} A = T_{AB}{}^C \nabla_C A$$
where the extra term in $T_{AB}{}^C$ from the new Lie derivative drops out because of the identity $(e^A A)e_A B = 0$ (and similarly for the $\nabla_A \ln \Phi^2$ term in $\tilde{T}_A$).

This means that the only kinds of covariant derivatives defined are:

$$\text{scalar gradient:} \quad \nabla_{\mathcal{A}} A$$
$$\text{"off-diagonal" gradient:} \quad \nabla_{\mathcal{A}} V_{\tilde{\mathcal{B}}}$$
$$\text{Lie derivative:} \quad V_{[1}^{\mathcal{B}} \nabla_{\mathcal{B}} V_{2]}^{\mathcal{A}} - \tfrac{1}{2} V_{[1\mathcal{B}} \nabla^{\mathcal{A}} V_{2]}^{\mathcal{B}} + V_1^{\mathcal{B}} V_2^{\mathcal{C}} T_{\mathcal{BC}}{}^{\mathcal{A}}$$
$$\text{divergence:} \quad \nabla_{\mathcal{A}} V^{\mathcal{A}}$$

(and $\mathcal{A} \leftrightarrow \tilde{\mathcal{A}}$) and combinations of these (since the covariant derivative is a linear operator). In the supersymmetric case there are also certain spinor derivatives:

$$g^{(b,\tilde{\beta})C} \nabla_{[C} V_{\alpha)}, \qquad including \quad \text{spinor curl:} \quad \nabla_{(\alpha} V_{\beta)}$$

$$\Pi_{\alpha}{}^{\beta}{}_{\gamma}{}^{\delta} g^{\gamma E} \nabla_{[E} V_{\delta)}$$

Another way to see the importance of this gradient and divergence is to note that the transformation laws of the vielbein and dilaton written with gauge parameters with tangent-space indices are

$$\delta \Phi^2 = \Phi^2 \nabla_A \lambda^A$$

$$(\delta e_{AM}) e^{BM} = -g^{BC} (\nabla_{[A} \lambda_{C)} - T_{AC}{}^D \lambda_D) + (\lambda_A{}^B - \lambda^C \omega_{CA}{}^B)$$

In the second transformation law the latter set of terms vanishes for appropriate values of the indices (the same for vanishing of $\lambda_A{}^B$ as for vanishing of $\omega_{CA}{}^B$): For example,

$$(\delta e_{\mathcal{A}}{}^M) e_{\tilde{\mathcal{B}}M} = -(\delta e_{\tilde{\mathcal{B}}}{}^M) e_{\mathcal{A}M} = -(\nabla_{[\mathcal{A}} \lambda_{\tilde{\mathcal{B}})} - T_{\mathcal{A}\tilde{\mathcal{B}}}{}^C \lambda_C)$$

In the conventional gauge these covariant derivatives become fully covariant with respect to the residual Lorentz gauge invariance: For the bosonic case,

$$\omega_{\tilde{a}b}{}^c = \overset{\circ}{\omega}_{ab}{}^c + \tfrac{1}{2} H_{ab}{}^c, \quad \omega_{a\tilde{b}}{}^{\tilde{c}} = \overset{\circ}{\omega}_{ab}{}^c - \tfrac{1}{2} H_{ab}{}^c$$

where $\overset{\circ}{\omega}$ is the usual matter-free Lorentz connection, and thus the torsion $T_{abc} = \pm H_{abc}$ is the usual left- or right-handed torsion of the string [18].

## 8. CURVATURES AND ACTIONS

In this section we consider mostly just the bosonic case. To construct curvature tensors (including field equations) and actions, we first consider the linear approximation:
$$e_A{}^M \equiv \langle e_A{}^M \rangle + h_A{}^B \langle e_B{}^M \rangle \quad \Rightarrow \quad e_A \approx d_A \equiv \langle e_A{}^M \rangle \partial_M$$

Because of the gauge transformations and constraint
$$\delta h_{AB} \approx \lambda_{AB} - d_{[A}\lambda_{B)}, \quad 0 = g_{A\tilde{B}} \approx h_{A\tilde{B}} + h_{\tilde{B}A}$$

we can consider without loss of generality only expressions involving $h_{\mathcal{A}\tilde{\mathcal{B}}}$ and the linearized dilaton:
$$\Phi^2 \equiv 1 + \phi$$
$$\delta\phi \approx d^A \lambda_A, \quad \delta h_{\mathcal{A}\tilde{\mathcal{B}}} \approx -d_{[\mathcal{A}}\lambda_{\tilde{\mathcal{B}})}$$

Then we find there are no objects first-order in derivatives which are invariant under these gauge transformations (torsions). In the supersymmetric case, the tangent-space gauge group is restricted, so there are torsions, as discussed in section 3. However, even under these unrestricted transformations there are some invariants second-order in derivatives (curvatures):

$$R \approx \Box\phi + d^{\mathcal{A}} d^{\tilde{\mathcal{B}}} h_{\mathcal{A}\tilde{\mathcal{B}}}$$
$$R_{\mathcal{A}\tilde{\mathcal{B}}} \approx \Box h_{\mathcal{A}\tilde{\mathcal{B}}} - d_{\mathcal{A}} d^{\mathcal{C}} h_{\mathcal{C}\tilde{\mathcal{B}}} + d_{\tilde{\mathcal{B}}} d^{\tilde{\mathcal{C}}} h_{\mathcal{A}\tilde{\mathcal{C}}} + d_{\mathcal{A}} d_{\tilde{\mathcal{B}}}\phi$$
$$R_{\mathcal{A}\mathcal{B}\tilde{\mathcal{C}}\tilde{\mathcal{D}}} \approx d_{[\mathcal{A}} d_{[\tilde{\mathcal{C}}} h_{\mathcal{B}]\tilde{\mathcal{D}}]}$$

$$\Box = d^{\mathcal{A}} d_{\mathcal{A}} = -d^{\tilde{\mathcal{A}}} d_{\tilde{\mathcal{A}}}$$

The analogous tensors in pure gravity are the Ricci scalar, traceless Ricci tensor, and Weyl tensor (which they contain). They also have a corresponding physical interpretation: $R = 0$ and $R_{\mathcal{A}\tilde{\mathcal{B}}} = 0$ are the field equations from varying $\Phi$ ($\phi$) and $e_A{}^M$ ($h_{\mathcal{A}\tilde{\mathcal{B}}}$), while $R_{\mathcal{A}\mathcal{B}\tilde{\mathcal{C}}\tilde{\mathcal{D}}}$ is the on-shell field strength. However, in this case they cannot be combined into a single curvature tensor because there are no such covariant traces. (This is similar to supergravity, where they fall into separate multiplets.) The simplest globally covariant gauge is

$$f_{\mathcal{A}} \equiv d^{\tilde{\mathcal{B}}} h_{\mathcal{A}\tilde{\mathcal{B}}} + \tfrac{1}{2} d_{\mathcal{A}}\phi = f_{\tilde{\mathcal{A}}} \equiv -d^{\mathcal{A}} h_{\mathcal{A}\tilde{\mathcal{B}}} + \tfrac{1}{2} d_{\tilde{\mathcal{B}}}\phi = 0$$

$$\Rightarrow \quad R_{\mathcal{A}\tilde{\mathcal{B}}} \approx \Box h_{\mathcal{A}\tilde{\mathcal{B}}}, \quad R \approx \tfrac{1}{2}\Box\phi$$

These tensors satisfy the linearized Bianchi identities

$$d_{[\mathcal{A}} R_{\mathcal{BC}]\tilde{\mathcal{D}}\tilde{\mathcal{E}}} \approx 0$$
$$d_{[\mathcal{A}} R_{\mathcal{B}]\tilde{\mathcal{C}}} \approx d^{\tilde{\mathcal{D}}} R_{\mathcal{AB}\tilde{\mathcal{C}}\tilde{\mathcal{D}}}$$
$$d_{\mathcal{A}} R \approx -d^{\tilde{\mathcal{B}}} R_{\mathcal{A}\tilde{\mathcal{B}}}$$

and similarly for $\mathcal{A} \leftrightarrow \tilde{\mathcal{A}}$.

The linearized gauge-invariant and gauge-fixed lagrangians are

$$-\int \delta L \approx \int R\delta\phi + R_{\mathcal{A}\tilde{\mathcal{B}}}\delta h^{\mathcal{A}\tilde{\mathcal{B}}}$$
$$\Rightarrow \quad L \approx -\tfrac{1}{2}\phi\Box\phi - \phi d^{\mathcal{A}} d^{\tilde{\mathcal{B}}} h_{\mathcal{A}\tilde{\mathcal{B}}} - \tfrac{1}{2}h^{\mathcal{A}\tilde{\mathcal{B}}}\Box h_{\mathcal{A}\tilde{\mathcal{B}}} - \tfrac{1}{2}(d^{\mathcal{A}} h_{\mathcal{A}\tilde{\mathcal{B}}})^2 + \tfrac{1}{2}(d^{\tilde{\mathcal{B}}} h_{\mathcal{A}\tilde{\mathcal{B}}})^2$$
$$\Rightarrow \quad L - \tfrac{1}{2}f_{\mathcal{A}}^2 + \tfrac{1}{2}f_{\tilde{\mathcal{A}}}^2 \approx -\tfrac{1}{4}\phi\Box\phi - \tfrac{1}{2}h^{\mathcal{A}\tilde{\mathcal{B}}}\Box h_{\mathcal{A}\tilde{\mathcal{B}}}$$

We now consider the fully nonlinear generalization. We can define naive covariant derivatives satisfying the torsion constraints, but we must demand that the undetermined pieces of the connection drop out of all true tensors. The naive curvatures, defined by replacing $f_{[AB)}{}^C$ with $F_{AB}{}^C$ in the usual definition to make them $\lambda^M$-covariant,

$$r_{(1)ABC}{}^D \equiv e_{[A}\omega_{B)C}{}^D + \omega_{[A|C}{}^E \omega_{|B)E}{}^D - F_{AB}{}^E \omega_{EC}{}^D$$

are $\lambda_A{}^B$-noncovariant as a result of this replacement,

$$\delta r_{(1)ABC}{}^D = (covariant) - \tfrac{1}{2}\omega_{EC}{}^D (e^E \lambda_{[A|}{}^F) g_{F|B)}$$

After a modification to simplify this transformation law

$$r_{(2)ABC}{}^D \equiv r_{(1)ABC}{}^D - \tfrac{1}{2}\omega_{EC}{}^D e^E g_{AB}$$
$$\Rightarrow \quad \delta r_{(2)ABC}{}^D = (covariant) - \omega_{EC}{}^D (e^E \lambda_A{}^F) g_{FB}$$

and a further redefinition to make the transformation more symmetric

$$r_{(3)ABC}{}^D \equiv r_{(2)ABC}{}^D - \tfrac{1}{2}\omega_{EC}{}^D \omega^E{}_{AB}$$
$$\Rightarrow \quad \delta r_{(3)A}{}^B{}_C{}^D = (covariant) - \tfrac{1}{2}\omega_{EC}{}^D e^E \lambda_A{}^B + \tfrac{1}{2}\omega_{EA}{}^B e^E \lambda_C{}^D$$

we can define a covariant curvature tensor (but still containing undefined connections)

$$R_{ABCD} \equiv \tfrac{1}{2}(r_{(3)ABCD} + r_{(3)CDAB})$$

Truly covariant tensors can be defined by taking appropriate traces, so that the remaining components of the connection are only those that can be defined by the above torsion constraints. It is sufficient to look for nonlinear generalizations of the curvatures obtained from the linearized analysis. Unfortunately, because the undefined components of the connections do not drop out of any of the above $r_{AB\tilde{C}\tilde{D}}$'s, there is no nonlinear form of this curvature. (In particular, because of the index symmetrization in $R_{ABCD}$, the desired linearized expression cancels.) This situation is similar (and perhaps related) to string field theory, where covariant, nonlinear expressions exist for field equations but not for on-shell field strengths. However, nonlinear forms of the other tensors do exist:

$$[\nabla_\mathcal{A}, \nabla_{\tilde{\mathcal{B}}}\} V^{\tilde{\mathcal{B}}} \equiv R_{\mathcal{A}\tilde{\mathcal{B}}} V^{\tilde{\mathcal{B}}}$$

$$\Rightarrow \quad R_{\mathcal{A}\tilde{\mathcal{B}}} = 2R_{\tilde{\mathcal{C}}\mathcal{A}\tilde{\mathcal{B}}}{}^{\tilde{\mathcal{C}}} = 2R_{\mathcal{C}\tilde{\mathcal{B}}\mathcal{A}}{}^{\mathcal{C}} = (e_{\tilde{\mathcal{B}}}\widetilde{F}_\mathcal{A} - F_{\tilde{\mathcal{B}}\mathcal{A}}{}^{\mathcal{C}}\widetilde{F}_\mathcal{C}) - (e_\mathcal{C} F_{\tilde{\mathcal{B}}\mathcal{A}}{}^{\mathcal{C}} - F_{\mathcal{C}\tilde{\mathcal{B}}}{}^{\tilde{\mathcal{E}}} F_{\tilde{\mathcal{E}}\mathcal{A}}{}^{\mathcal{C}})$$

where we have used the torsion constraints and $F$ Bianchi identities. We also have

$$R = -\tfrac{1}{2} R_{\mathcal{AB}}{}^{\mathcal{AB}} = \tfrac{1}{2} R_{\tilde{\mathcal{A}}\tilde{\mathcal{B}}}{}^{\tilde{\mathcal{A}}\tilde{\mathcal{B}}}$$
$$= e_\mathcal{A} \widetilde{F}^\mathcal{A} + \tfrac{1}{2}\widetilde{F}_\mathcal{A}{}^2 + \tfrac{1}{2} e_\mathcal{A} e_\mathcal{B} g^{\mathcal{AB}} - \tfrac{1}{2} F_{\mathcal{AB}\tilde{\mathcal{C}}}{}^2 - \tfrac{1}{432} F_{[\mathcal{ABC})}{}^2 + \tfrac{1}{8}(e^\mathcal{A} g^{\mathcal{BC}})(e_\mathcal{B} g_{\mathcal{AC}})$$

where the last expression equals minus the same expression with left- and right-handed indices switched.

Since the full nonlinear action must be homogeneous of order two in $\Phi$ (since it is the only density around), and $R = 0$ is the only duality-covariant field equation that can result from varying the action with respect to $\Phi^2$ (this requires $R$ be a scalar, since $\Phi^2$ is a density), it follows that

$$R \equiv -\frac{\delta S}{\delta \Phi^2} \quad \Rightarrow \quad S = -\int d^D x \; \Phi^2 R$$

(We can also write a cosmological term $\int d^D x \; \Phi^2$.) From the general variation of the action

$$-\delta S = \int R \; \delta\Phi^2 + \Phi^2 R^{\mathcal{A}\tilde{\mathcal{B}}} e_{\tilde{\mathcal{B}}}{}^M \delta e_{\mathcal{A}M}$$

we find that invariance of the action $\delta S = 0$ under gauge transformations requires

$$\nabla^{\tilde{\mathcal{B}}} R_{\mathcal{A}\tilde{\mathcal{B}}} + \nabla_\mathcal{A} R = \nabla^\mathcal{B} R_{\mathcal{B}\tilde{\mathcal{A}}} - \nabla_{\tilde{\mathcal{A}}} R = 0$$

which are just the Bianchi identities (as in ordinary gravity).

For $R_{\mathcal{A}\tilde{\mathcal{B}}}$, the torsion constraint $T_{\mathcal{A}\tilde{\mathcal{B}}}{}^C = 0$ is necessary to make undefined connections drop out. On the other hand, $R$ can be expressed in terms of the torsion and the vielbein (and its derivatives) alone, which allows the bosonic action to be written in first-order form. The identity between the $-\frac{1}{2}R_{\mathcal{AB}}{}^{\mathcal{AB}}$ and $\frac{1}{2}R_{\tilde{\mathcal{A}}\tilde{\mathcal{B}}}{}^{\tilde{\mathcal{A}}\tilde{\mathcal{B}}}$ forms holds only for the second-order formalisms (i.e. after imposing the torsion constraints); before imposing the torsion constraints,

$$-R_{AB}{}^{AB} = \nabla_A \widetilde{T}^A + \tfrac{1}{2}\widetilde{T}_A{}^2 + \tfrac{1}{2}\nabla_a \nabla_B g^{AB} - \tfrac{1}{432}T_{[ABC]}{}^2 - \tfrac{1}{96}(\nabla_{(A}g_{BC]})^2$$

The connection-independent pieces of this identity are a Bianchi identity of the $F$'s. The two forms of $R$ give two different first-order formalisms for the theory. For example,

$$\begin{aligned}-R_{\mathcal{AB}}{}^{\mathcal{AB}} &= \tfrac{1}{12}(\tfrac{1}{2}\omega^{[\mathcal{ABC})2} + \tfrac{1}{3}\omega^{[\mathcal{ABC})}F_{[\mathcal{ABC})}) + \tfrac{1}{2}(\tfrac{1}{2}\omega_{\mathcal{B}}{}^{[\mathcal{AB})2} - 2e_{\mathcal{A}}\omega_{\mathcal{B}}{}^{[\mathcal{AB})} - \omega_{\mathcal{B}}{}^{[\mathcal{AB})}e^{\mathcal{C}}g_{\mathcal{AC}}) \\ &+ \tfrac{1}{12}(\tfrac{1}{2}\omega^{(\mathcal{ABC}]2} + \tfrac{1}{2}\omega^{(\mathcal{ABC}]}e_{(\mathcal{A}}g_{\mathcal{BC}]}) - \tfrac{1}{2}(\tfrac{1}{2}\omega_{\mathcal{B}}{}^{(\mathcal{AB}]2} + \omega_{\mathcal{B}}{}^{(\mathcal{AB}]}e^{\mathcal{C}}g_{\mathcal{AC}}) \\ &+ (\tfrac{1}{2}\omega^{\tilde{\mathcal{A}}\mathcal{BC}2} + \omega^{\tilde{\mathcal{A}}\mathcal{BC}}F_{\tilde{\mathcal{A}}\mathcal{BC}})\end{aligned}$$

(The term $-e_{\mathcal{A}}\omega_{\mathcal{B}}{}^{[\mathcal{AB})}$ produces a term $\widetilde{F}_{\mathcal{A}}\omega_{\mathcal{B}}{}^{[\mathcal{AB})}$ upon integration by parts in the lagrangian $-\Phi^2 R$.) Thus, varying the connections in the first-order form

$$S = \int d^D x \; \Phi^2 \tfrac{1}{4}(R_{\mathcal{AB}}{}^{\mathcal{AB}} - R_{\tilde{\mathcal{A}}\tilde{\mathcal{B}}}{}^{\tilde{\mathcal{A}}\tilde{\mathcal{B}}})$$

gives the correct second-order form, as well as determining almost all the connections we have already found.

In the supersymmetric case the form of the off-shell field strengths (appearing in the field equations) follows from the usual ones for the scalar and vector multiplets, at least for D=3 and 4 (and to some extent, 6): The generalization of $R$ comes from varying the action with respect to the dilaton superfield (scalar multiplet). Therefore, in D=3 it is a real scalar superfield, while in D=4 it is a chiral scalar superfield; in D=6 the form depends on the formulation of the scalar multiplet. These superfields already appear in pure supergravity (since the scalar multiplet appears as a superscale compensator there also). The generalization of $R_{\mathcal{A}\tilde{\mathcal{B}}}$ is found by taking the field equations of the vector multiplets, which carry an index $\hat{b}$, and generalizing that index $\hat{b} \to \tilde{\mathcal{B}}$ to preserve the local tangent-space symmetry. (So $R_{a\hat{b}}$, which represents the field equations from varying the vector gauge fields $A_{a\hat{b}}$, becomes $R_{a\tilde{\mathcal{B}}} = R_{\mathcal{A}\tilde{\mathcal{B}}}$ in the bosonic case.) Thus in D=3 the vector multiplet gauge superfield $A_{\alpha\hat{a}}$ leads to $R_{\alpha\tilde{\mathcal{B}}}$, while in D=4 the scalar prepotential $U_{\hat{a}}$ yields $R_{\tilde{\mathcal{A}}}$, and in D=6 the scalar, isotriplet prepotential $V_{(ij)}$ gives $R_{(ij)\tilde{\mathcal{A}}}$. For example, in D=4 $R_{\tilde{a}}$ is the usual second field equation for pure supergravity, but it now has absorbed the tensor multiplet in the string gauge for superscale transformations. The bosonic field strength $R_{a\tilde{\mathcal{B}}}$ appears at order $\theta\bar\theta$ in $R_{\tilde{\mathcal{B}}}$. This structure reflects the fact that the closed-string spectrum is the direct product of open-string spectra: The closed-string superfield is the direct product of a single left-handed vector-multiplet superfield (as a function of the left-handed spinor coordinate $\theta$) with a right-handed bosonic multiplet $\psi_{\tilde{\mathcal{A}}}$ consisting of a vector $\psi_{\tilde{a}}$ and scalars $\psi_{\hat{a}}$. (The scalar multiplet, representing the superscale compensator, is the direct product of a left-handed super Yang-Mills-ghost scalar multiplet with a right-handed bosonic Yang-Mills-ghost scalar [9].)

## 9. SUMMARY

Finally we outline the fundamental results: This approach to certain field theories is derived from the string by beginning with the oscillator algebra

$$[Z_M(1), Z_N(2)\} = i\delta'(2-1)\eta_{MN}$$

which defines a global symmetry (duality) acting on these superindices, as well as a local symmetry with group elements $e^{-i\Lambda}$ generated by $Z_M$,

$$\Lambda = \int \lambda^M Z_M$$

The gauge invariance of the gauge invariance

$$\delta\lambda^M = \partial^M \lambda \quad \Rightarrow \quad \delta\Lambda = 0$$

follows from the fundamental identities

$$A' = Z^M \partial_M A, \quad (\partial^M A)(\partial_M B) = \partial^M \partial_M A = 0$$

which reflect the triviality of winding-mode dependence (but in a weaker way). The group algebra defines a new Lie derivative

$$[\Lambda_1, \Lambda_2] = -i\Lambda_{[1,2]} \quad \Rightarrow \quad \lambda^M_{[1,2]} = \lambda^N_{[1} \partial_N \lambda^M_{2]} - \tfrac{1}{2}\lambda^N_{[1}\partial^M \lambda_{2]N}$$

Background fields are coupled by covariantizing $Z_M$,

$$\Pi_A = e_A{}^M Z_M, \quad g_{AB} = e_A{}^M e_B{}^N \eta_{MN}$$

Their algebra defines $\lambda^M$-covariant objects $F$:

$$[\Pi_A(1), \Pi_B(2)\} = i\delta'(2-1)\tfrac{1}{2}[g_{AB}(1) + g_{AB}(2)] + i\delta F_{AB}{}^C \Pi_C$$

The definition of the first-class (Virasoro) constraints restricts the tangent-space metric and gauge group:

$$g_{\mathcal{A}\tilde{\mathcal{B}}} = 0$$

$$\Rightarrow \quad L_+ = \tfrac{1}{2} g^{\mathcal{A}\mathcal{B}} \Pi_\mathcal{B} \Pi_\mathcal{A}, \quad L_- = \tfrac{1}{2} g^{\tilde{\mathcal{A}}\tilde{\mathcal{B}}} \Pi_{\tilde{\mathcal{B}}} \Pi_{\tilde{\mathcal{A}}}; \quad L_+ + L_- = \tfrac{1}{2} Z^M Z_M$$

The resulting gauge transformations of the background fields are:

$$\delta \Pi_A = [-i\Lambda, \Pi_A] + \lambda_A{}^B \Pi_B, \quad \lambda_\mathcal{A}{}^{\tilde{\mathcal{B}}} = \lambda_{\tilde{\mathcal{A}}}{}^\mathcal{B} = 0$$

$$\Rightarrow \quad \delta e_A{}^M = (\lambda^N \partial_N e_A{}^M + e_{AN} \partial^{[M} \lambda^{N]}) + \lambda_A{}^B e_B{}^M, \quad \delta g_{AB} = \lambda^M \partial_M g_{AB} + \lambda_{(AB]}$$

The dilaton is introduced as a density to allow actions:

$$\delta \Phi^2 = \partial_M (\lambda^M \Phi^2)$$

In the most general case (supersymmetry) the indices take the values $M = (\mathbf{M}, {}^\mathbf{M}, \hat{m})$, $A = (\mathcal{A}, \tilde{\mathcal{A}}) = (\alpha, a, \tilde{\alpha}; \tilde{a}, \hat{a})$. The identification of $\Pi_\alpha$ as the second-class constraints produces further restrictions on the tangent-space:

$$g_{\alpha\beta} = 0, \quad \lambda_\alpha{}^b = \lambda_\alpha{}^{\tilde{\beta}} = \Pi_\alpha{}^\beta{}_\delta{}^\gamma \lambda_\gamma{}^\delta = 0$$

As a result, the connection-independent torsions are

$$F_{\alpha\beta}{}^{\tilde{\mathcal{C}}}, \ F_{\alpha\beta}{}^{c,\tilde{\gamma}}, \ \Pi_{\alpha\beta}{}^\gamma{}_\delta{}^{\epsilon\zeta} F_{\epsilon\zeta}{}^\delta, \ F_{\alpha\tilde{\mathcal{B}}}{}^{c,\tilde{\gamma}}, \ \Pi_\alpha{}^\gamma{}_\delta{}^\epsilon F_{\epsilon\tilde{\mathcal{B}}}{}^\delta$$

The requirement that the constraint algebra takes the usual form constrains these torsions:

$$F_{\alpha\beta}{}^{\tilde{\mathcal{C}}} = F_{\alpha\beta\gamma} = F_{\alpha\tilde{\mathcal{B}}}{}^{c,\tilde{\gamma}} = 0, \quad F_{\alpha\beta}{}^{c,\tilde{\gamma}} = \gamma^{\check{d}}_{\alpha\beta} F_{\check{d}}{}^{c,\tilde{\gamma}}$$

To preserve the last (representation-preserving) constraint, $\lambda_\alpha{}^\beta$ must be constrained to Lorentz, scale, and internal symmetries. This constraint also allows definition of the $\kappa$-symmetry generators. The Bianchi identities show that the remaining connection-independent torsions are also constrained:

$$\Pi_{\alpha\beta}{}^\gamma{}_\delta{}^{\epsilon\zeta} F_{\epsilon\zeta}{}^\delta = \Pi_\alpha{}^\gamma{}_\delta{}^\epsilon F_{\epsilon\tilde{\mathcal{B}}}{}^\delta = 0$$

If the tangent-space group is partially fixed to leave just the usual Lorentz and internal symmetries, the residual fields can be identified as the usual (super)fields. The resulting form of the constraints is in the string gauge with respect to (super)scale transformations, obtained from the usual background formalism for the string.

In D=3 the torsion constraints can be solved off shell for $e_A{}^M$ in terms of $e_\alpha{}^M$. In D=4 the representation-preserving constraints further determine the latter (and thus $\Pi_\alpha$) as

$$\Pi_\alpha = A_\alpha{}^\mu e^W Z_\mu e^{-W}, \quad \Pi_{\dot\alpha} = A_{\dot\alpha}{}^{\dot\mu} e^{-\overline{W}} Z_{\dot\mu} e^{\overline{W}}, \quad W = \int W^M Z_M$$

$$e_\alpha = A_\alpha{}^\mu e^w \partial_\mu e^{-w}, \quad w = W^M i \partial_M$$

in analogy to pure supergravity in ordinary superspace. Also as ordinary superspace, this solution generalizes the gauge group:

$$e^{W'} = e^{-i\Lambda} e^W e^{i\Xi}, \quad \Xi = \int \xi^M Z_M, \quad \partial_\mu \xi^M = 0 \; except \; \partial_\mu \xi^\nu \neq 0$$

plus the analog of the gauge invariance of the gauge invariance

$$W'^M = W^M + \partial^M \zeta$$

The new transformations replace the old when we work with the $\Lambda$-invariant combination

$$e^U \equiv e^{\overline{W}} e^W$$

Although we would like a fully general covariant derivative

$$\nabla_A = e_A + \omega_{A\mathcal{B}}{}^\mathcal{C} G_\mathcal{C}{}^\mathcal{B} + \omega_{A\tilde{\mathcal{B}}}{}^{\tilde{\mathcal{C}}} G_{\tilde{\mathcal{C}}}{}^{\tilde{\mathcal{B}}}$$

the new Lie derivative and corresponding gauge transformations modify the torsion to

$$T_{AB}{}^C = F_{AB}{}^C + (\omega_{[AB)}{}^C + \tfrac{1}{2} \omega^C{}_{[AB)})$$

so that the connection constraints

$$\omega\text{-dependent } T_{AB}{}^C = \nabla_A g_{BC} = 0$$

$$\Rightarrow \quad \omega_{A\tilde{\mathcal{B}}}{}^{\tilde{\mathcal{C}}} = -F_{A\tilde{\mathcal{B}}}{}^{\tilde{\mathcal{C}}}, \quad \omega_{A(\mathcal{BC})} = -e_A g_{\mathcal{BC}}, \quad \omega_{[ABC)} = -\tfrac{1}{3} F_{[ABC)}$$

do not determine all the connections. Integration by parts for the measure $\Phi^2$ defines the new torsion and connection constraint

$$\tilde{T}_A \equiv \Phi^2 \overleftarrow{\nabla}_A \Phi^{-2} = 0$$

$$\Rightarrow \quad \omega_{BA}{}^B = -\tilde{F}_A \equiv -\Phi^2 \overleftarrow{e}_A \Phi^{-2}$$

which determines another piece of the connection.

The identity $\partial^M \partial_M = 0$ makes traces a fundamental part of the definition of curvatures as well as torsions:

$$R_{\mathcal{A}\tilde{\mathcal{B}}} = 2R_{\tilde{\mathcal{C}}\mathcal{A}\tilde{\mathcal{B}}}{}^{\tilde{\mathcal{C}}} = 2R_{\mathcal{C}\tilde{\mathcal{B}}\mathcal{A}}{}^{\mathcal{C}} = (e_{\tilde{\mathcal{B}}}\widetilde{F}_\mathcal{A} - F_{\tilde{\mathcal{B}}\mathcal{A}}{}^{\mathcal{C}}\widetilde{F}_\mathcal{C}) - (e_\mathcal{C} F_{\tilde{\mathcal{B}}\mathcal{A}}{}^{\mathcal{C}} - F_{\mathcal{C}\tilde{\mathcal{B}}}{}^{\tilde{\mathcal{E}}} F_{\tilde{\mathcal{E}}\mathcal{A}}{}^{\mathcal{C}})$$

$$R = -\tfrac{1}{2}R_{\mathcal{A}\mathcal{B}}{}^{\mathcal{A}\mathcal{B}} = \tfrac{1}{2}R_{\tilde{\mathcal{A}}\tilde{\mathcal{B}}}{}^{\tilde{\mathcal{A}}\tilde{\mathcal{B}}}$$
$$= e_\mathcal{A}\widetilde{F}^\mathcal{A} + \tfrac{1}{2}\widetilde{F}_\mathcal{A}{}^2 + \tfrac{1}{2}e_\mathcal{A}e_\mathcal{B}g^{\mathcal{A}\mathcal{B}} - \tfrac{1}{2}F_{\mathcal{A}\mathcal{B}\tilde{\mathcal{C}}}{}^2 - \tfrac{1}{432}F_{[\mathcal{A}\mathcal{B}\mathcal{C}]}{}^2 + \tfrac{1}{8}(e^\mathcal{A}g^{\mathcal{B}\mathcal{C}})(e_\mathcal{B}g_{\mathcal{A}\mathcal{C}})$$

These curvatures satisfy the Bianchi identities

$$\nabla^{\tilde{\mathcal{B}}} R_{\mathcal{A}\tilde{\mathcal{B}}} + \nabla_\mathcal{A} R = \nabla^\mathcal{B} R_{\mathcal{B}\tilde{\mathcal{A}}} - \nabla_{\tilde{\mathcal{A}}} R = 0$$

In the bosonic case the action is simply

$$S = -\int d^D x \; \Phi^2 R$$

and the field equations $R = R_{\mathcal{A}\tilde{\mathcal{B}}} = 0$ come from the variation

$$-\delta S = \int R \; \delta\Phi^2 + \Phi^2 R^{\mathcal{A}\tilde{\mathcal{B}}} e_{\tilde{\mathcal{B}}}{}^M \delta e_{\mathcal{A}M}$$

## ACKNOWLEDGMENTS

I thank Nathan Berkovits, Jim Gates, and Martin Roček for discussions.